\documentclass{aa}
\usepackage{graphicx}
\begin{document}
%
%
\title{Using Cepheids to determine the galactic abundance gradient.II.
Towards the galactic center
\thanks{Based on spectra collected at AAO - Australia}}
\titlerunning{Galactic abundance gradient}
\author{S.M. Andrievsky,
\inst{1,2}\,
D. Bersier,
\inst{3}\,
V.V. Kovtyukh,
\inst{2,4}\,
R.E. Luck,
\inst{5}\,\\
W.J. Maciel,
\inst{1}\,
J.R.D. L\'epine,
\inst{1}\,
Yu.V. Beletsky
\inst{2,4}}
\authorrunning{Andrievsky et al.}
\offprints{S.M. Andrievsky}
\institute{
Instituto Astron\^ {o}mico e Geof\' \i sico, Universidade de S\~{a}o
Paulo, Av. Miguel Stefano, 4200, S\~ao Paulo SP, Brazil\\
email:
sergei@andromeda.iagusp.usp.br
\and
Department of Astronomy, Odessa State University,
Shevchenko Park, 65014, Odessa, Ukraine\\
email:
scan@deneb.odessa.ua; val@deneb.odessa.ua
\and
Harvard-Smithsonian Center for Astrophysics, 60 Garden Street,
MS 16, Cambridge, MA 02138, USA\\
email: dbersier@cfa.harvard.edu
\and
Odessa Astronomical Observatory and Isaac Newton Institute of Chile, Odessa Branch,
Ukraine
\and
Department of Astronomy, Case Western Reserve
University, 10900 Euclid Avenue, Cleveland, OH 44106-7215\\
email: luck@fafnir.astr.cwru.edu}
\date{Received ; accepted }
\abstract
{Based on spectra obtained at the Anglo-Australian Observatory,
we present a discussion of the metallicity of the galactic disc
derived using Cepheids at galactocentric distances 4-6 kpc.
Our new results together with previous gradient determination (Paper I)
show that the overall abundance distribution within the galactocentric
distances 4--11~kpc cannot by represented by a single gradient value.
The distribution is more likely bimodal: it is flatter in the solar
neighbourhood with a small gradient, and steepens towards the galactic
center. The steepening begins at a distance of about 6.6 kpc.
\keywords{Stars: abundances--stars: Cepheids--Galaxy: abundances--Galaxy:
evolution}}
\maketitle

\section{Introduction}
In our previous work (Andrievsky et al. 2002, hereafter Paper I) results on elemental abundance
distributions in the galactic disc based on 236 high-resolution spectra of 77 classical Cepheids in
the solar neighborhood (galactocentric distances from 6 kpc to approximately 10.5 kpc) were
reported. We found that among the 25 studied chemical elements, those from carbon to yttrium
show small negative gradients, while heavier species produce near-to-zero gradients.
Typical gradient values for iron-group elements were found to be equal to
$\approx -0.03$ dex~kpc$^{-1}$.

In order to extend our previous study and to check the behavior of the elemental distribution
towards the galactic center we have observed several Cepheids with galactocentric distances
between 4-6 kpc. In this work we present the results from these stars and discuss them together
with the data from Paper I.

\section{Observations}

The spectra of the program stars (see Table 1) have been obtained on June
2nd 2001 at the Anglo-Australian Telescope with the University College of
London Echelle Spectrograph (UCLES). The detector is an MIT/Lincoln Lab CCD
with 15 micron pixels (2048 x 4096 pixels). With a 31 line~mm$^{-1}$ grating,
the resolving power is approximately 80000. The signal-to-noise ratio for all
spectra is greater than 100, and the total coverage is 5000-9000 \AA. The
spectra were processed with the help of the DECH20 package (Galazutdinov, 1992).

\begin{table}
\caption[]{Program stars and details of the observations}
\begin{tabular}{rrcrcc}
Star & V & UT & Exp. (s) & JD 2452063. &$\phi$ \\
\hline
\object{VY Sgr}  &11.51& 14:43:52 & 1800 & .1242  & 0.394\\
\object{UZ Sct}  &11.31& 15:15:51 & 1800 & .1464  & 0.455\\
\object{AV Sgr}  &11.39& 14:11:58 & 1800 & .1021  & 0.982\\
\object{V340 Ara}&10.16& 13:43:09 &  900 & .0768  & 0.167\\
\object{KQ Sco}  & 9.81& 13:59:43 &  600 & .0866  & 0.940\\
\hline
\end{tabular}
\end{table}

\section{Parameters of program stars and elemental abundances}

The methods applied for the determination of the atmosphere parameters and elemental
abundances are the same as described in detail in Paper I, and will not be repeated here. In
Tables 2 and 3 the derived parameters of the program stars and their elemental abundances
are given. Note that in particular, for the iron abundance, the standard deviation of the
abundances is about 0.1.  Given the number of iron lines utilized (44 - 102) the standard error of
the mean is about 0.01.
Following the scheme adopted in Paper I, in the gradient plots we have assigned weight 3 for all
the program stars except KQ Sco. The lines in the spectrum of this star show some asymmetry.
This factor limits the number of possible equivalent width measurements and may produce larger
errors in the analysis. A weight of 1 was assigned for this star.

\begin{table}
\caption[]{Atmosphere parameters for program stars}
\begin{tabular}{rcccccc}
\hline
Star &$\phi$&T$_{\rm eff}$,\,K& $\log g$ &V$_{\rm t}$,\, km~s$^{-1}$ \\
\hline
VY Sgr  & 0.394& 5144 & 1.25 & 3.30 \\
UZ Sct  & 0.455& 5127 & 1.50 & 3.30 \\
AV Sgr  & 0.982& 5875 & 1.75 & 4.90 \\
V340 Ara& 0.167& 5472 & 1.50 & 4.00 \\
KQ Sco  & 0.940& 5058 & 1.10 & 5.70 \\
\hline
\end{tabular}
\end{table}

\begin{table*}
\caption[]{Elemental abundances}
\tiny
\begin{tabular}{rrrrrrrrrrrrrrrr}
\hline
\hline
\noalign{\smallskip}
\multicolumn{1}{c}{}&\multicolumn{3}{c}{VY Sgr}& \multicolumn{3}{c}{UZ Sct}&
\multicolumn{3}{c}{AV Sgr}& \multicolumn{3}{c}{V340 Ara}& \multicolumn{3}{c}
{KQ Sco}\\
\noalign{\smallskip}
\hline
Ion &[M/H]&$\sigma$&N&[M/H]&$\sigma$&N&[M/H]&$\sigma$&N&[M/H]&$\sigma$&N&
[M/H]&$\sigma$&N\\
\hline
C I & 0.01&0.23& 3 & 0.04&0.15& 3 & 0.21&0.13&14& 0.20&0.11& 8 &--0.02& -- & 1 \\
N I & 0.50& -- & 1 & 0.70& -- & 1 & 0.82&0.11& 2& 1.00&0.12& 2 & -- & -- & -- \\
O I & 0.18&0.19& 2 & 0.49&0.21& 5 & 0.36&0.05& 2& 0.07&0.20& 2 & 0.21& -- & 1 \\
Na I & 0.66& -- & 1 & 0.76& -- & 1 & 0.58&0.06& 2& 0.56& -- & 1 & -- & -- & -- \\
Mg I & -- & -- & -- & -- & -- & -- &--0.12& -- & 1& -- & -- &-- & -- & -- & -- \\
Al I & 0.42&0.05& 2 & 0.44& -- & 1 & 0.68& -- & 1& 0.35&0.07& 2 & 0.42& -- & 1 \\
Si I & 0.30&0.07& 13 & 0.33&0.13& 14 & 0.35&0.07&13& 0.35&0.10&18 & 0.21& 0.08& 6\\
S I  & 0.56&0.15& 3 & 0.63&0.16& 2 & 0.39&0.02& 3& 0.51&0.28& 4 & 0.33& 0.19& 3 \\
Ca I & 0.23&0.13& 3 & 0.35&0.21& 3 & 0.13&0.01& 2& 0.21&0.23& 3 & 0.39& 0.16& 2 \\
Sc I & 0.31& -- & 1 & 0.36& -- & 1 & -- & -- &--& -- & -- &-- & -- & -- & -- \\
Sc II& --  & -- &-- & -- & -- & -- & 0.32& -- & 1& 0.33& -- & 1 & -- & -- & -- \\
Ti I & 0.22&0.11& 14 & 0.24&0.09& 12 & 0.40&0.16& 7& 0.28&0.20&13 & 0.22& 0.11& 4 \\
Ti II& 0.51& -- & 1 & 0.41& -- & 1 & 0.22& -- & 1& 0.49& -- & 1 & 0.12& -- & 1 \\
V I  & 0.20&0.09& 12 & 0.24&0.04& 7 & 0.40&0.08& 9& 0.21&0.12&10 & 0.21& 0.12& 6 \\
V II & -- & -- & -- & 0.17& -- & 1 & 0.26&0.05& 2& 0.22&0.12& 3 & 0.05& -- & 1 \\
Cr I & 0.22&0.18& 9 & 0.25&0.20& 5 & 0.22&0.10& 3& 0.09& -- & 1 & 0.18& 0.11& 3 \\
Cr II& -- & -- & -- & 0.22& -- & 1 & 0.49& -- & 1& -- & -- &-- & 0.60& -- & 1 \\
Mn I & 0.22&0.00& 2 & 0.11&0.15& 2 & 0.12&0.21& 3& 0.25&0.35& 2 &--0.09& -- & 1 \\
Fe I & 0.26&0.08&102 & 0.33&0.09& 95 & 0.34&0.12&83& 0.31&0.08&85 & 0.16& 0.07&44 \\
Fe II& 0.27&0.11& 12 & 0.33&0.14& 16 & 0.35&0.09&18& 0.34&0.06&14 & 0.15& 0.04& 4\\
Co I & 0.24&0.13& 12 & 0.19&0.12& 11 & 0.11& -- & 1& 0.15&0.13& 3 & 0.12& 0.19& 6 \\
Ni I & 0.33&0.10& 34 & 0.31&0.08& 15 & 0.38&0.12&20& 0.39&0.09&25 & 0.13& 0.06&14 \\
Cu I & 0.24& -- & 1 & 0.43& -- & 1 & 0.41& -- & 1& 0.29& -- & 1 &--0.06& -- & 1 \\
Zn I & 0.78& -- & 1 & 0.78& -- & 1 & 0.61& -- & 1& 0.46& -- & 1 & -- & -- & -- \\
Sr I & 0.24& -- & 1 & -- & -- & -- & -- & -- &--& -- & -- &-- & -- & -- & -- \\
Y II & --& -- & -- & 0.43&0.20& 2 & 0.13& -- & 1& 0.34&0.05& 2 & -- & -- & -- \\
Zr II&--0.04& -- & 1 & 0.05& -- & 1 & 0.14& -- & 1& 0.20& -- & 1 & -- & -- & -- \\
La II& 0.25& -- & 1 & 0.29& -- & 1 & 0.21& -- & 1& 0.19& -- & 1 & 0.05& -- & 1 \\
Ce II&--0.01&0.13& 3 & 0.09&0.21& 2 & 0.00& -- & 1&--0.01&0.24& 3 &--0.31& -- & 1 \\
Nd II&--0.07&0.13& 3 &--0.01&0.12& 5 & 0.10&0.36& 2& 0.05&0.28& 4 & 0.13& 0.37& 2 \\
Eu II& 0.20&0.20& 2 & 0.25&0.18& 2 & 0.39& -- & 1& 0.29& -- & 1 & 0.02& -- & 1 \\
Gd II& 0.18& -- & 1 & 0.38& -- & 1 & 0.35& -- & 1& 0.21& -- & 1 & -- & -- & -- \\
\hline
\end{tabular}
\end{table*}

\section{Galactocentric distances}
Galactocentric distances for program Cepheids were calculated using the same procedure as
described in Paper I. Photometric data were taken from the catalogue of Fernie et al. (1995).
Results are given in Table 4 together with other useful data.

\begin{table*}
\caption[]{Some physical and positional characteristics of program Cepheids}
\scriptsize
\begin{tabular}{cccccccccc}
\hline
Star & P, d &$<$B-V$>$ & E(B-V) &M$_{\rm v}$& d, pc & l & b & R$_{\rm G}$, kpc&
$<$[Fe/H]$>$\\
\hline
VY Sgr  & 13.5572& 1.941& 1.283&--4.43& 2187  & 10.13 &--1.08& 5.76& +0.26 \\
UZ Sct  & 14.7442& 1.784& 1.071&--4.53& 2867  & 19.13 &--1.50& 5.28& +0.33 \\
AV Sgr  & 15.415 & 1.999& 1.267&--4.58& 2250  & 7.53  &--0.59& 5.68& +0.34 \\
V340 Ara& 20.809 & 1.539& 0.574&--4.94& 4321  & 335.19&--3.75& 4.38& +0.31 \\
KQ Sco  & 28.6896& 1.934& 0.896&--5.33& 2623  & 340.39&--0.75& 5.50& +0.16 \\
\hline
\end{tabular}
\end{table*}

\section{Discussion}

In order to cover galactocentric distances from 4 kpc to 11 kpc we combined the
data from the present study with those obtained in Paper I. The results are
shown in Figs. 1-5 . An obvious increase of the abundances towards the galactic
center is seen for many elements. For all elements except possibly the heaviest
ones shown in Fig. 5, a rather flat distribution in the solar neighborhood begins
to steepen at approximately 6.6 kpc. This is particularly seen for iron, our most
reliable abundance. From our data the overall distribution (e.g., d[Fe/H]/dR) over
the baseline considered is difficult to represent with a single gradient value.
More likely the distribution is bimodal, as shown in Fig. 6, which contains the
same data as Fig. 1, but where we formally considered two possible zones separated
by a boundary at approximately 6.6 kpc. Formally, this could be represented by a
[Fe/H] gradient of $-0.14$ dex~kpc$^{-1}$ for $4<$ R$_{\rm G}$(kpc)$<6.6$ and of
$-0.02$ dex~kpc$^{-1}$ $6.6<$ R$_{\rm G}$(kpc)$<12$.

The steepening can be also visually traced in Figs. 2-4 for some other chemical
elements. However, this is not the case for heavy species (Fig. 5) for which in
Paper I we found near-to-zero gradients. The present results confirm that finding.

We do not believe that a simple extrapolation of our results into the inner
zone nearer the center of the Galaxy (R$_{\rm G} << 4$ kpc) is warranted. It
is possible that the chemical evolution of the central galactic region differs
from the rest of the galactic disc, and that close to the galactic center the
abundance distribution may become flatter. There are some indications from the
recent studies of Ram\' \i rez et al. (\cite{ret00}) and Carr, Sellgren \&
Balachandran (\cite{caret00}) that some relatively young M supergiant stars
at the very galactic center possess solar metallicities.

\begin{figure}
 \includegraphics[height=6.0cm]{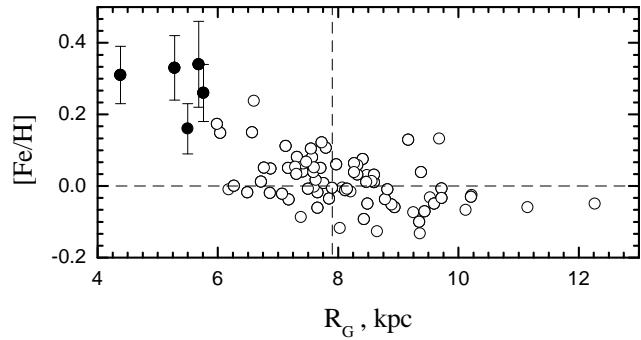}
\caption[]{The radial distribution of the iron abundance. $Open~circles$ -
data from Paper I, $black~circles$ - present results. 2-$\sigma$ interval
is indicated for the stars from present study.}
\end{figure}

\begin{figure*}
 \centering
 \includegraphics[height=21.5cm]{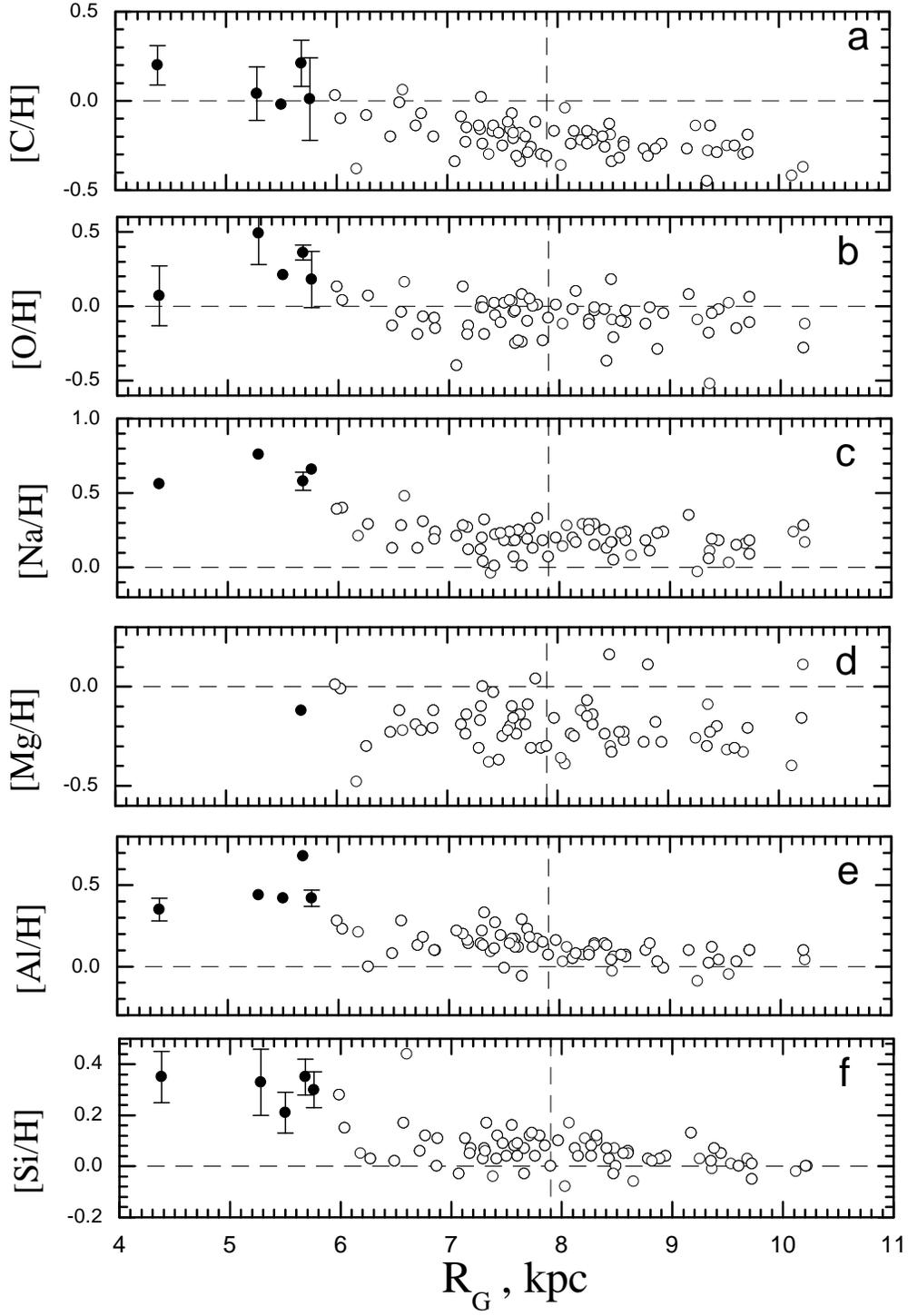}
\caption[]{Same as Fig. 1, but for elements C--Si}
\end{figure*}

\begin{figure*}
 \centering
 \includegraphics[height=21.5cm]{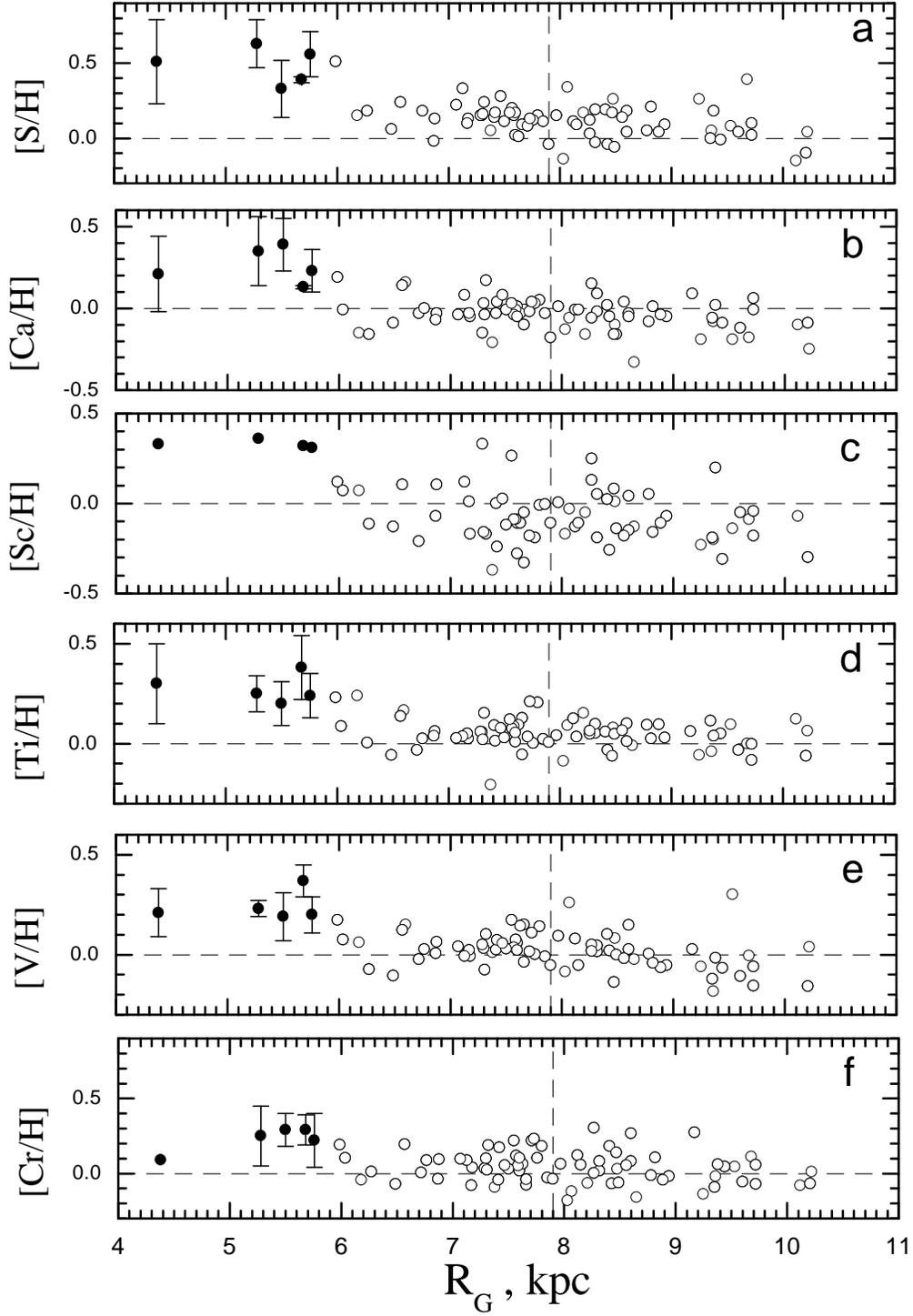}
\caption[]{Same as Fig. 1, but for elements S--Cr}
\end{figure*}

\begin{figure*}
 \centering
 \includegraphics[height=21.5cm]{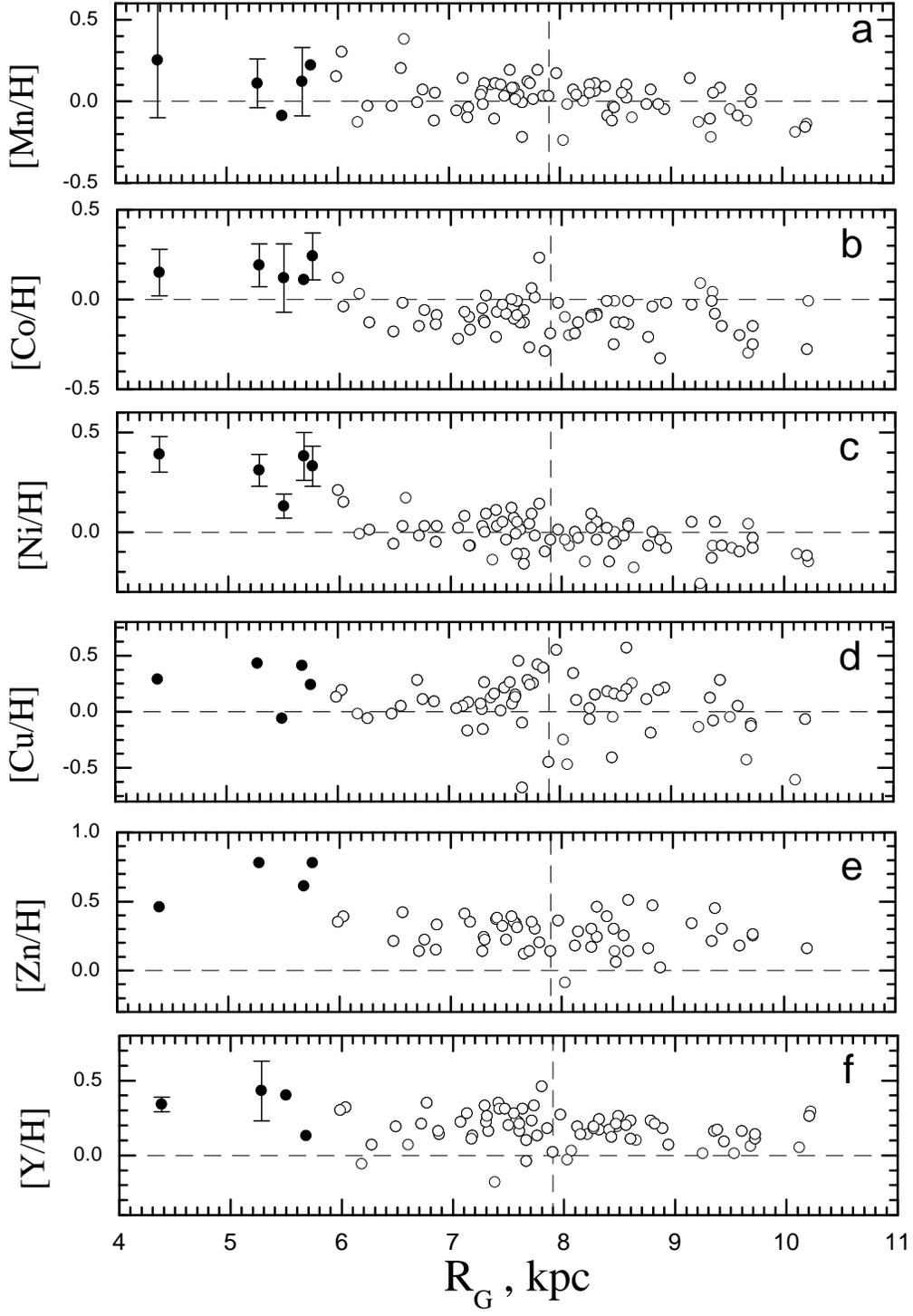}
\caption[]{Same as Fig. 1, but for elements Mn--Y}
\end{figure*}

\begin{figure*}
 \centering
 \includegraphics[height=21.5cm]{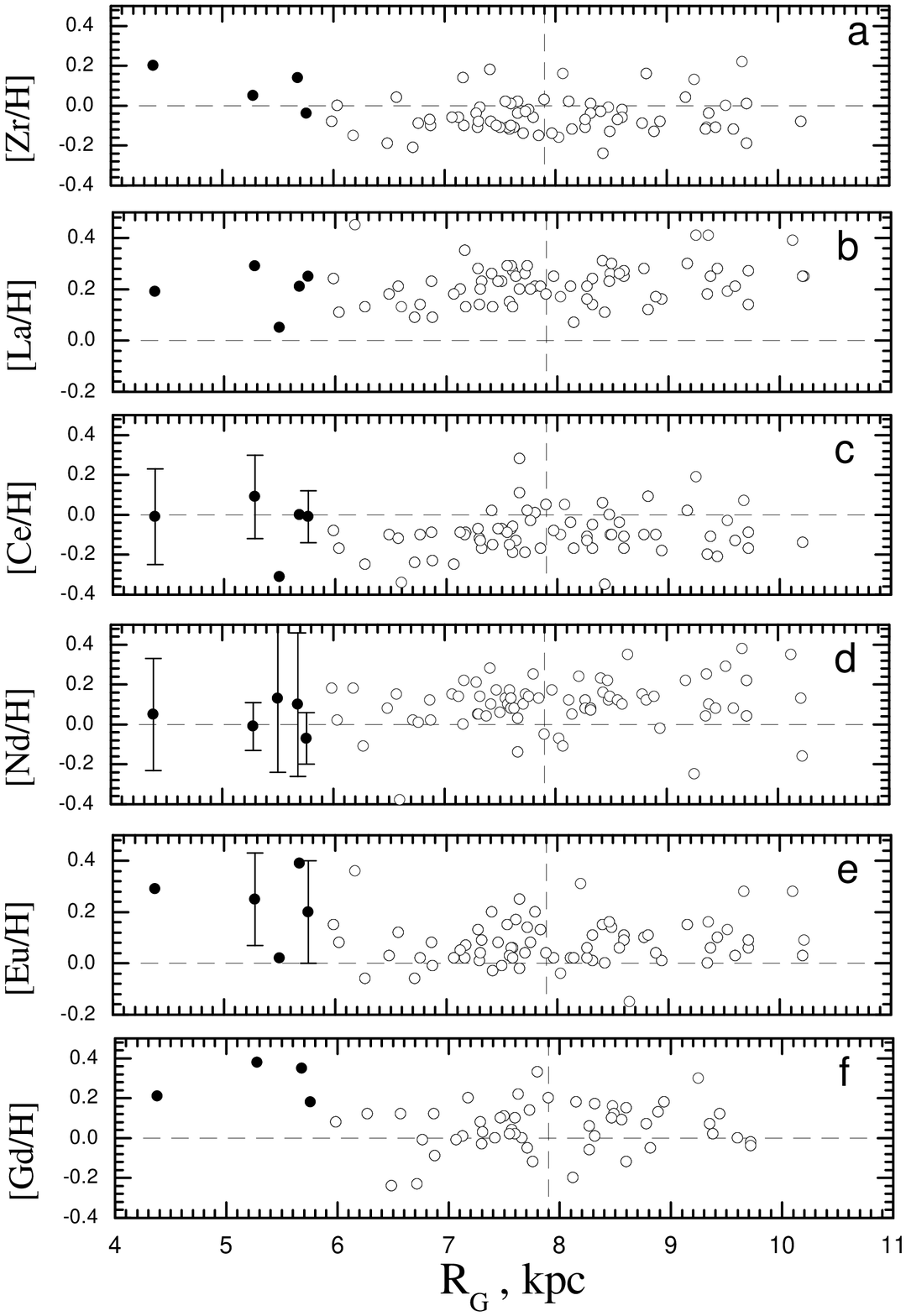}
\caption[]{Same as Fig. 1, but for elements Zr--Gd}
\end{figure*}

Below we briefly discuss a comparison between the metallicity distribution in
the inner and middle part of galactic disc obtained from Cepheids with the
most recent observational and theoretical results.

Recent determinations of the [Fe/H] gradient in the galactic disc have been
on the basis of open cluster stars (see Friel 1995, 1999, Phelps 2000). Although
these objects may have a wider age span than Cepheids, it is interesting to
compare those gradients with our results, especially in view of the supposition
that abundance gradients have not varied significantly in the last few Gyrs
(Maciel \& Quireza 1999, Maciel \& Costa 2002). The review by Friel (1995) indicates
a gradient of $-0.091$ dex~kpc$^{-1}$, while a more recent presentation of
data by the same group indicates a flatter value of $-0.06$ dex~kpc$^{-1}$
(Friel 1999, Phelps 2000). Gradients similar to the steeper value above were
also suggested by Carraro, Ng, Portinari (1998) and Bragaglia et al. (2000),
and the results reported by Twarog, Ashman \& Antony-Twarog (1997) also show
some slope variation at about 10 kpc from the galactic center.

It can be seen that the above gradient results are intermediate between our
steeper "inner" gradient and the flatter "outer" gradient shown in Fig. 6.
In fact, if we roughly take an average of both regions we get d[Fe/H]/dR around
$-0.08$ dex~kpc$^{-1}$, which is close to the average gradient value for
open clusters.

In Paper I we discussed the connection between the character of the
elemental distribution in the solar neigbourhood and galactic bar which
may induce the radial flows in the disc, and thus may produce a significant
homogenization. Another observable phenomenon, which should be also caused
by the bar, is a metallicity increase in the inner part of the disc.
As was shown by Martinet \& Fridli (1997), a rather young and strong bar
produces two distinct gradients in the disc, one steep in the inner part,
and another shallow in the outer region.

According to the model of Portinari \& Chiosi (\cite{pch00}), a rotating
bar sweeps the gas out from its co-rotation radius to the outer Lindblad
resonance (located at approximately 5 kpc from the center), so that some
local increase in the metallicity is expected in this region (see Fig. 14
for oxygen distribution from above mentioned paper). Qualitatively, this
is confirmed by our observational results, but the detected oxygen
overabundance at 4--6 kpc is higher than that predicted by the model.
It should be noted that Portinari \& Chiosi (\cite{pch00}) presumed in
their model that bar has negligible influence on the disc beyond the
outer Linblad resonance. They also used in the numerical simulation
the velocity of the gas flows of about 0.1-1.0 km~s$^{-1}$, although
according to Stark (1984) and Stark \& Brand (1989) the radial flows
in the solar neighbourhood have a velocity of about 4 km~s$^{-1}$.

Chiappini, Matteucci \& Romano (\cite{cmr01}) presented a chemical evolution
model assuming two accretion episodes in the Galaxy formation, but no radial
flows.
We overlay their "model A" gradient on our data in Fig. 6. In the region from
4--6 kpc their model fits the data adequately but outward of 6 kpc the model
predicts abundances significantly in excess of our observed values. If one
rescales their gradient to [Fe/H] = 0 at the solar radius, then the predicted
abundances in the inner region are significantly lower than the observed values.
Note that as their model does not consider the possibility of homogenization
in the disc from the radial flows triggered by a galactic bar, this factor might
be responsible for some vertical shift between our observational data and the
model prediction for the solar vicinity.

Summarizing, both an accretion-based galactic evolution model without
the radial flows, and a model including the radial flows in the
disc, produce a local increase of the abundance distribution at galactocentric
distances of about 4--6 kpc. Such an increase was also detected in our
observational study. Nevertheless, disagreement exists in the absolute
abundances predicted by the models and those found from our observations.
This disagreements may arise either from ignoring the radial flows, or
from inadequate characteristics of the flows adopted in the model.

\begin{figure}
\resizebox{\hsize}{!}{\includegraphics{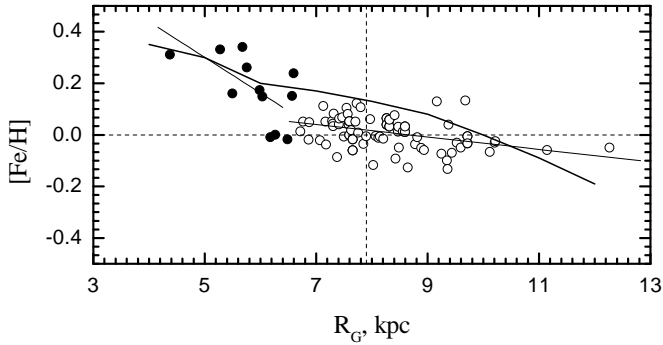}} \caption[]{The
radial distribution of the iron abundance for two zones.
$Open~circles$ and $black~circles$ represent the Cepheids
separated at 6.6 kpc. Theoretical prediction of "inside-out" model
A (Chiappini, Matteucci \& Romano (\cite{cmr01}) - $thick~line$.}
\end{figure}

\section{Conclusion}

We supplemented our previous data on elemental abundance distributions in the
solar neighborhood (Paper I) with new determinations based on Cepheids at
distances of 4-6 kpc. Our new results together with previous gradient
determinations (Paper I) show that the abundance distribution over the
galactocentric distances 4--11~kpc cannot by represented by a single gradient
value. More likely, the distribution is bimodal: it is flatter in the solar
neighborhood with a small gradient, and becomes steeper towards the galactic
center. The steepening begins at the distance about 6.5 kpc.

\newpage

\begin{acknowledgements}
SMA would like to express his gratitude to FAPESP for the visiting professor
fellowship (No. 2000/06587-3) and to Instituto Astron\^ {o}mico e Geof\'\i sico,
Universidade de S\~{a}o Paulo for providing facility support during a
productive stay in Brazil. The authors thank S. Ryder and S. Lee for acquiring
the spectra.
\end{acknowledgements}

\end{document}